\documentclass[utf8]{frontiersSCNS} 

\usepackage{url,lineno,microtype,subcaption}
\usepackage[onehalfspacing]{setspace}

\usepackage{bm}
\usepackage{epsfig} 
\usepackage{mathptmx} 
\usepackage{times} 
\usepackage{amsmath} 
\usepackage{amssymb}  
\usepackage{algorithm}
\usepackage{algorithmic}

\DeclareMathOperator*{\GEM}{GEM}

\DeclareMathOperator*{\DP}{DP}

\makeatletter
  \newcommand{\figcaption}[1]{\def\@captype{figure}\caption{#1}}
  \newcommand{\tblcaption}[1]{\def\@captype{table}\caption{#1}}
\makeatother

\def\keyFont{\fontsize{8}{11}\helveticabold }
\def\firstAuthorLast{Nakashima {et~al.}} 
\def\Authors{Ryo Nakashima\,$^{1,*}$, Ryo Ozaki\,$^{1}$ and Tadahiro Taniguchi\,$^{1}$}
\def\Address{$^{1}$Emergent Systems Laboratory, College of Information Science and Engineering, Ritsumeikan University, Shiga, Japan}
\def\corrAuthor{Tadahiro Taniguchi}
\def\corrEmail{taniguchi@em.ci.ritsumei.ac.jp}

\begin{document}
\onecolumn
\firstpage{1}

\title[Double Articulation Analyzer and Neural Network with Parametric Bias]{Unsupervised Phoneme and Word Discovery from Multiple Speakers using Double Articulation Analyzer and neural network with Parametric Bias} 

\author[\firstAuthorLast ]{\Authors} 
\address{} 
\correspondance{} 

\extraAuth{}

\maketitle

\begin{abstract}
This paper describes a new unsupervised machine learning method for simultaneous phoneme and word discovery from multiple speakers. 
Human infants can acquire knowledge of phonemes and words from interactions with his/her mother as well as with others surrounding him/her.
From a computational perspective, phoneme and word discovery from multiple speakers is a more challenging problem than that from one speaker because the speech signals from different speakers exhibit different acoustic features.
This paper proposes an unsupervised phoneme and word discovery method that simultaneously uses nonparametric Bayesian double articulation analyzer (NPB-DAA) and deep sparse autoencoder with parametric bias in hidden layer (DSAE-PBHL). 
We assume that an infant can recognize and distinguish speakers based on certain other features, e.g., visual face recognition. 
DSAE-PBHL is aimed to be able to subtract speaker-dependent acoustic features and extract speaker-independent features. 
An experiment demonstrated that DSAE-PBHL can subtract distributed representations of acoustic signals, enabling extraction based on the types of phonemes rather than on the speakers. Another experiment demonstrated that a combination of NPB-DAA and DSAE-PB outperformed the available methods in phoneme and word discovery tasks involving speech signals with Japanese vowel sequences from multiple speakers.
\tiny
\keyFont{ \section{Keywords:} word discovery, phoneme discovery, parametric bias, Bayesian model, neural network} 
\end{abstract}

\section{Introduction}\label{sec:1}
Infants can discover phonemes and words from speech signals uttered by individuals surrounding them without transcribed data, i.e., labeled data, in a manner that differs from most automatic speech recognition systems (ASRs) developed recently~\citep{Saffran1996,Saffran1996a}.
This study is aimed at creating a machine learning method that can discover phonemes and words from unlabeled data for developing a constructive model of language acquisition by human infants and for leveraging the large amount of unlabeled data spoken by multiple speakers in the context of developmental robotics~\citep{Taniguchi2016SER}.

Most available ASRs are trained with transcribed data that need to be prepared separately from the learning process~\citep{Sugiura2015,Kawahara2000,Dahl2012}. By using certain supervised learning methods and certain model architectures, an ASR can be developed with a very large amount of transcribed speech data corpus, i.e., a set of pairs of text data and acoustic data. 
However, human infants can discover phoneme and words through their developmental process.
They do not need transcribed data. Moreover, they discover phonemes and words at a time when they have not developed the capability to read text data. This evidence implies that infants discover phonemes and words in an unsupervised manner, i.e., from his/her sensor--motor information.

It is widely established that eight-month-old children can also infer chunks of phones, i.e., word-like unit, from the distribution of acoustic signals~\citep{Saffran1996a}. Caregivers generally utter a sequence of words rather than an isolated word in their infant-directed speech ~\citep{Aslin1996}.
Therefore, word segmentation and discovery is essential for language acquisition.
Saffran et al. described that human infants use three types of cues for word segmentation: prosodic, distributional, and co-occurrence~\citep{Saffran1996a}. In this study, we focus on distributional cues. 
Saffran et al. reported that eight-month-old infants can also perform word segmentation from continuous speech by using solely distributional cues~\citep{Saffran1996}. Thiessen et al. reported that distributional cues appear to be used by human infants by the age of seven months~\citep{Thiessen2003}. This is earlier than for other cues.

However, the computational models that discover phonemes and words from human speech signals have not been completely explored in the fields of developmental robotics and natural language or speech processing~\citep{Lee2012,Lee2013a,Lee2015,Kamper2015,Taniguchi2016,Taniguchi2016b}. 
The unsupervised word segmentation problem has been studied for a long time~\citep{Brent1999,Venkataraman2001,Goldwater2008,Goldwater2009,Mochihashi2009,Johnson2009,Chen2014,Magistry2012,Sakti2011,takeda2017unsupervised}. 
However, these models are established to be incapable of providing satisfactory results if they are applied to phoneme sequences recognized by a phoneme recognizer, which usually involve a lot of phoneme recognition errors. This is because they do not consider phoneme recognition errors or posterior distribution of phonemes, i.e., probabilistic modeling of phoneme recognition.
Neubig et al. extended the sampling procedure proposed by Mochihashi to handle word lattices that can be obtained from an ASR system~\citep{Neubig2012}. However, the improvement was limited, and they did not consider phoneme acquisition. It was indicated that feedback information from segmented words is essential in phonetic category acquisition~\citep{Feldman2013}. 
Subsequent to these studies, several others have been conducted to develop unsupervised phoneme and word discovery~\citep{Lee2015,Kamper2015,Taniguchi2016,Taniguchi2016b}.
This type of research is mostly equivalent to the development of unsupervised learning of speech recognition system, which transforms speech signals to sequences of words. The development of an unsupervised machine learning method that can discover words and phonemes
is also important to provide fresh insight into developmental studies from a computational perspective. In this study, we employ Nonparametric Bayesian double articulation analyzer (NPB-DAA)~\citep{Taniguchi2016}. 

The double articulation structure existing in spoken language is a characteristic structural feature of human language~\citep{Chandler2002}. When we develop an unsupervised machine learning method based on probabilistic generative models, i.e., Bayesian approach, it is critical to clarify our assumption about the latent structure embedded in observation data. The double articulation structure is a two-layer hierarchical structure; i.e., a sentence is generated by stochastic transitions between words, a word corresponds to a deterministic sequence of phonemes, and a phoneme exhibits similar acoustic features. This double articulation structure is universal for languages.
NPB-DAA was developed to enable a robot to obtain knowledge of phonemes and words in an unsupervised manner even if the robot does not know the number of phonemes and words, lists of phonemes and words and transcription of the speech signals.
Taniguchi et al. introduced deep sparse autoencoder (DSAE) to improve the performance of NPB-DAA; they demonstrated that it also outperformed a conventional off-the-shelf ASR system trained using transcribed data~\citep{Taniguchi2016b}. Although it did not outperform the state-of-the-art deep learning-based ASR system, the performance was remarkable; this is considering that the main research purpose of developing NPB-DAA with DSAE was to develop an unsupervised phoneme and word discovery system that can be regarded as a computational explanation of the process of human language acquisition, rather than to develop a high-performance ASR system.

However, the experiments conducted in~\citep{Taniguchi2016,Taniguchi2016b} used speech data obtained from only one speaker. The NPB-DAA with DSAE did not assume learning environments where a robot modelling a human infant learns phonemes and words from multiple speakers. 
Human infants do not acquire knowledge of phonemes and words from his/her mother alone; they do so also from multiple speakers surrounding him/her. Therefore, the direct application of NPB-DAA with DSAE to the multi-speaker scenario is highly likely to be ineffective. How to extend NPB-DAA with DSAE to the multi-speaker scenario is the research question of this paper.

\begin{figure}[t]
  \centering
    \includegraphics[width=\linewidth]{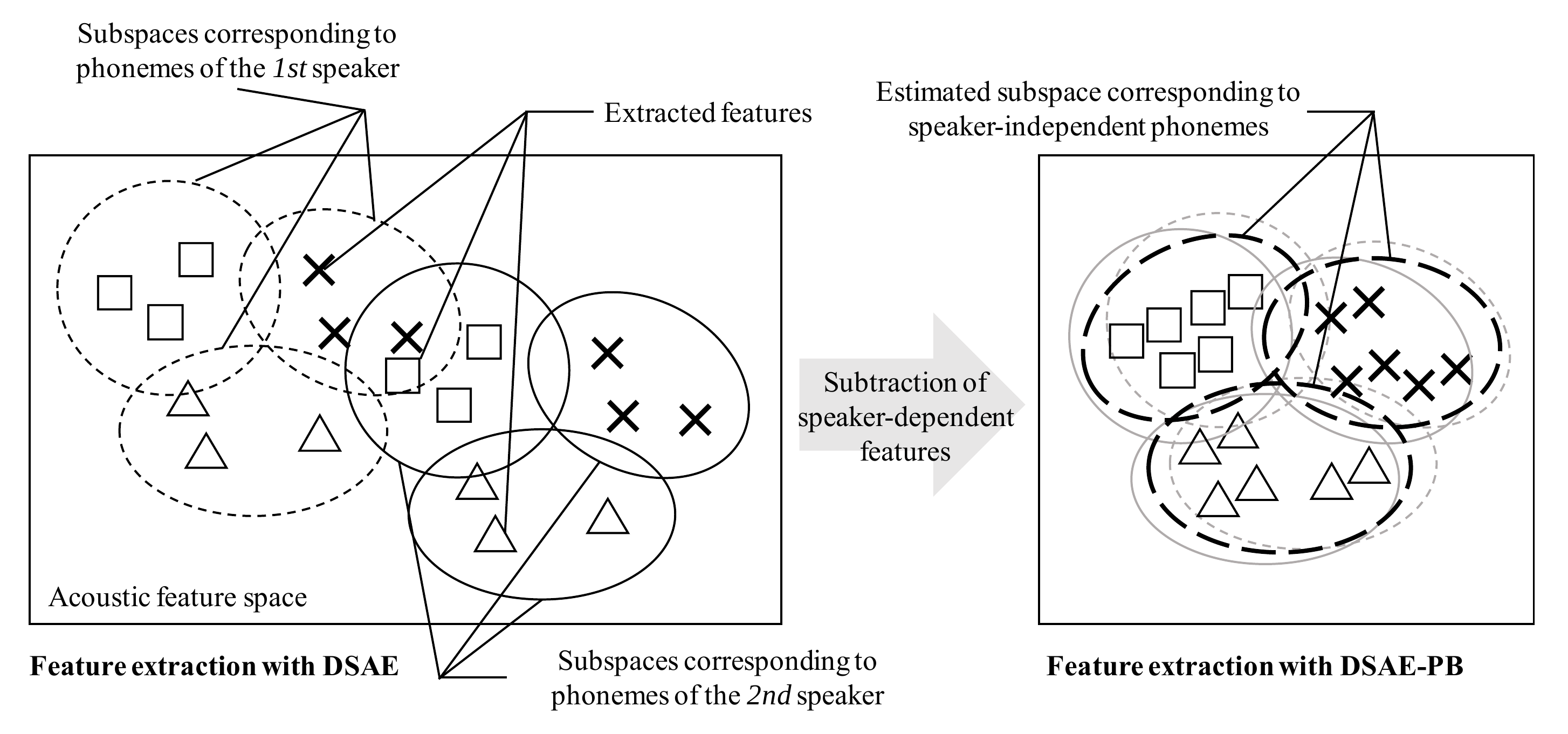}
  \caption{Schematic view of speaker-dependent and speaker-independent acoustic features and clustering result of the,}
  \label{fig:speaker-dependent-features}
\end{figure}

In the studies of unsupervised phoneme and word discovery, learning from speech signals obtained from multiple speakers have been recognized as challenging~\citep{dunbar2017zero,kamper2017embedded}.
To explain the essential challenge of the problem, let us consider an example of the discrimination of ``a'' from ``i.''
Figure~\ref{fig:speaker-dependent-features} provides a schematic view of the explanation that follows. 
Fundamentally, the phoneme discovery problem can be regarded as a type of clustering problem.
A machine learning method for unsupervised phoneme and word discovery should be capable of identifying clusters of ``a'' and ``i,'' and distinguishing them. If the acoustic feature distributions of ``a'' and ``i'' are sufficiently different, a proper unsupervised machine learning method can form two clusters, i.e., acoustic categories. For example, DSAE can form reasonable feature representations, and NPB-DAA can simultaneously categorize phonemes and words. If explicit feature representations are formed, a standard clustering method, e.g., Gaussian mixture model, can also perform phoneme discovery to a certain extent. 
However, in a multi-speaker setting, acoustic feature distribution of each phoneme can differ depending on the speakers. That is, ``a'' from the first speaker and ``a'' from the second speaker exhibit different feature distributions in the feature space. 
The direct application of a clustering method on the data tends to form different clusters, i.e., phoneme categories, for ``a'' from the first and second speakers. To enable a robot to acquire phonemes and words from speech signals obtained from multiple speakers, it needs to omit, cancel, or subtract speaker-dependent information from the observed speech signals. 
In Figure~\ref{fig:speaker-dependent-features}, the speaker-dependent features are subtracted, and the speaker-independent features are extracted. 
If speaker-independent feature representations can be formed similarly, the proposed clustering method, e.g., NPB-DAA, is likely to identify phonemes from the extracted features.

How to omit, cancel, or subtract speaker-dependent information is a crucial challenge in unsupervised phoneme and word discovery from multiple speakers. Conventional studies on ASR, which can use transcribed data, adopt an approach that omits the difference between multiple speakers by using transcribed data. Although ``a'' from speakers A and B exhibit different distributions, by using label data, the pattern recognition system can learn that both the distributions should be mapped to a label ``a.''  In the scenario of supervised learning, 
deep learning-based speech recognition systems adopt these types of approaches by exploiting a considerable amount of labeled data and the flexibility of neural networks~\citep{chan2016listen,chiu2018state,amodei2016deep,hannun2014deep}.
This approach was not suitable for this study because the research question is different; through this study, we intended investigate unsupervised phoneme and word discovery. 

The system should not use transcription. Instead of transcription, This study focused on information of speaker index, i.e., ``who is speaking now,'' to subtract speaker-dependent acoustic features.
It is widely established that infants can distinguish individuals around them in their early developmental stage. Therefore, the assumption that they can sense ``who is speaking now,'' i.e., speaker index, is reasonable from the developmental perspective. 

To apply speaker index and subtract speaker-dependent information from acoustic features, we employed the concept of parametric bias in the study of neural networks. 
Neural networks have been demonstrated to exhibit rich representation learning capability and widely used for a decade~\citep{Le2011,Krizhevsky2012,Liu2014a,Bengio2009,Hinton06}. 
In the context of developmental robotics, Tani and Ogata et al. proposed and explored recurrent neural network with parametric bias~\citep{tani2004self,yokoya2007experience,ogata2007two}. Parametric bias is an additional input that can function as a {\it gray} switch that can modify the function of the neural network.
In our study, the speaker index is provided as an input of parametric bias.
Moreover, the characteristic of neural networks wherein they encode independent feature information in each neuron if it is trained under suitable conditions is called disentanglement. 
The property of disentanglement has been attracting attention in recent studies~\citep{chen2016infogan,higgins2017beta,bengio}. 
The arithmetic manipulability rooting on this characteristic of the neural network has been gaining attention. 
It was demonstrated that Word2Vec, i.e., skip-gram for word embedding, can predict the representation vector of ``Paris'' by subtracting the vector of ``Japan'' from that of ``Tokyo'' and adding that of ``France''~\citep{mikolov2013distributed,mikolov2013efficient}. 
Considering these concepts, we propose DSAE with parametric bias in hidden layer (DSAE-PBHL) to subtract speaker-dependent information.

\begin{figure}[t]
  \centering
    \includegraphics[width=\linewidth]{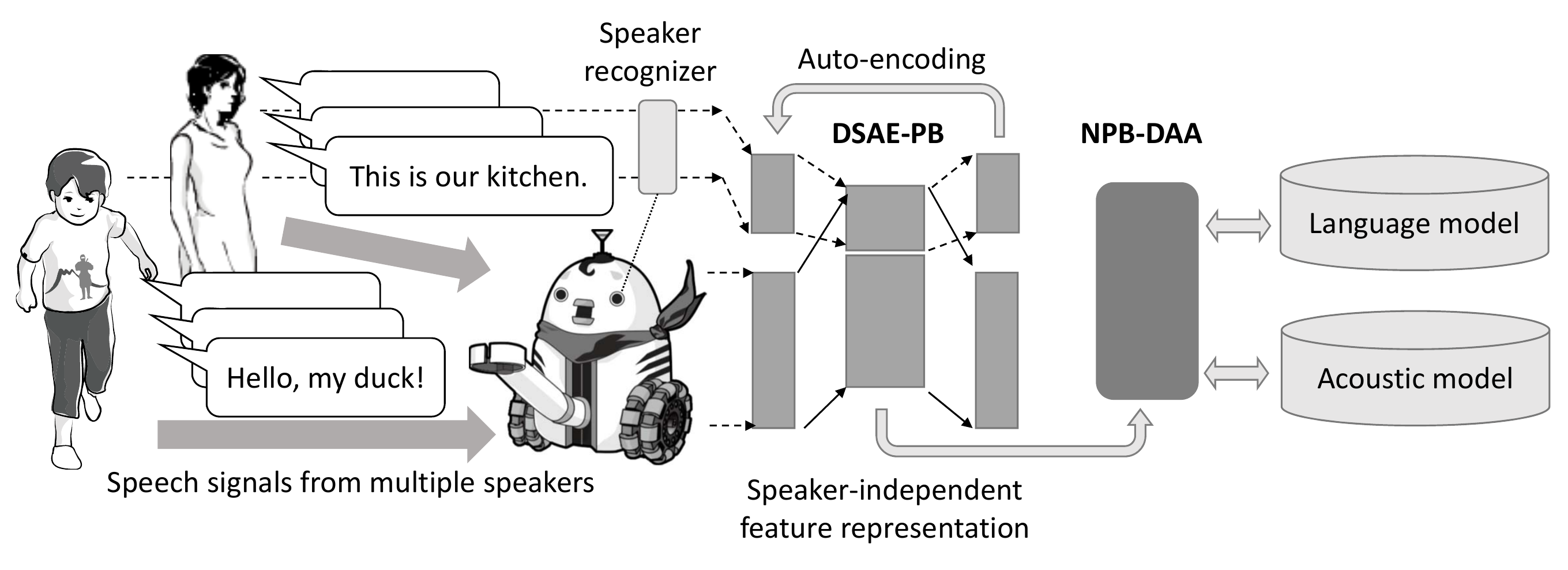}
  \caption{Overview of proposed method, NPB-DAA with DSAE-PBHL. First, a robot observes spoken utterances with speaker indexes using a speaker recognition method, e.g., face recognition. DSAE-PB, which accepts speaker-dependent features and speaker index as input, extracts speaker-independent feature representations and passes them to NPB-DAA. NPB-DAA segments the feature sequences and identifies words and phonemes, i.e., language and acoustic models, in an unsupervised manner.}
  \label{fig:overview}
\end{figure}

The overview of our approach, unsupervised phoneme and word discovery using  NPB-DAA with DSAE-PB, is schematically depicted in Figure~\ref{fig:overview}.  First, a robot observes spoken utterances with speaker indexes using a speaker recognition method, e.g., face recognition. DSAE-PB, which accepts speaker-dependent features and speaker index as input, extracts speaker-independent feature representations and passes them to NPB-DAA. NPB-DAA segments the feature sequences and identifies words and phonemes, i.e., language and acoustic models, in an unsupervised manner.

Our contribution is that we propose an unsupervised learning method that can identify words and phonemes directly from speech signals uttered by multiple speakers. The method based on NPB-DAA and DSAE-PBHL is an unsupervised learning method except for the use of an index of a speaker, which is assumed to be estimated by the robot, i.e., a model of a human infant.  

The remainder of this paper is organized as follows: 
Section~\ref{sec:2} briefly describes the proposed method, a combination of NPB-DAA and DSAE-PBHL. 
Section~\ref{sec:3} describes two experiments that evaluate the effectiveness of the proposed method using actual sequential Japanese vowel speech signals. Section~\ref{sec:4} concludes this paper.

\section{Methods}\label{sec:2}
The proposed method consists of NPB-DAA and DSAE-PBHL (see Figure~\ref{fig:overview}). First, we briefly introduce NPB-DAA~\citep{Taniguchi2016}. Secondly, we describe DSAE-PBHL after introducing DSAE~\citep{Ng2011,Liu2015iv,Taniguchi2016b}.

\subsection{NPB-DAA}

Hierarchical Dirichlet process hidden language model (HDP-HLM) is a probabilistic generative model that models  double articulation structure (i.e., two-layer hierarchy, a characteristic of human spoken language)~\citep{Taniguchi2016}. 
Mathematically, HDP-HLM is a natural extension of hierarchical Dirichlet process hidden semi-Markov model (HDP-HSMM), which is a type of generalization of hidden Markov model~\citep{Johnson2013}. 
NPB-DAA is the name of an unsupervised learning method for phoneme and word discovery based on HDP-HLM. 

\begin{figure*}[t]
  \centering
  \includegraphics[width = 0.5\linewidth]{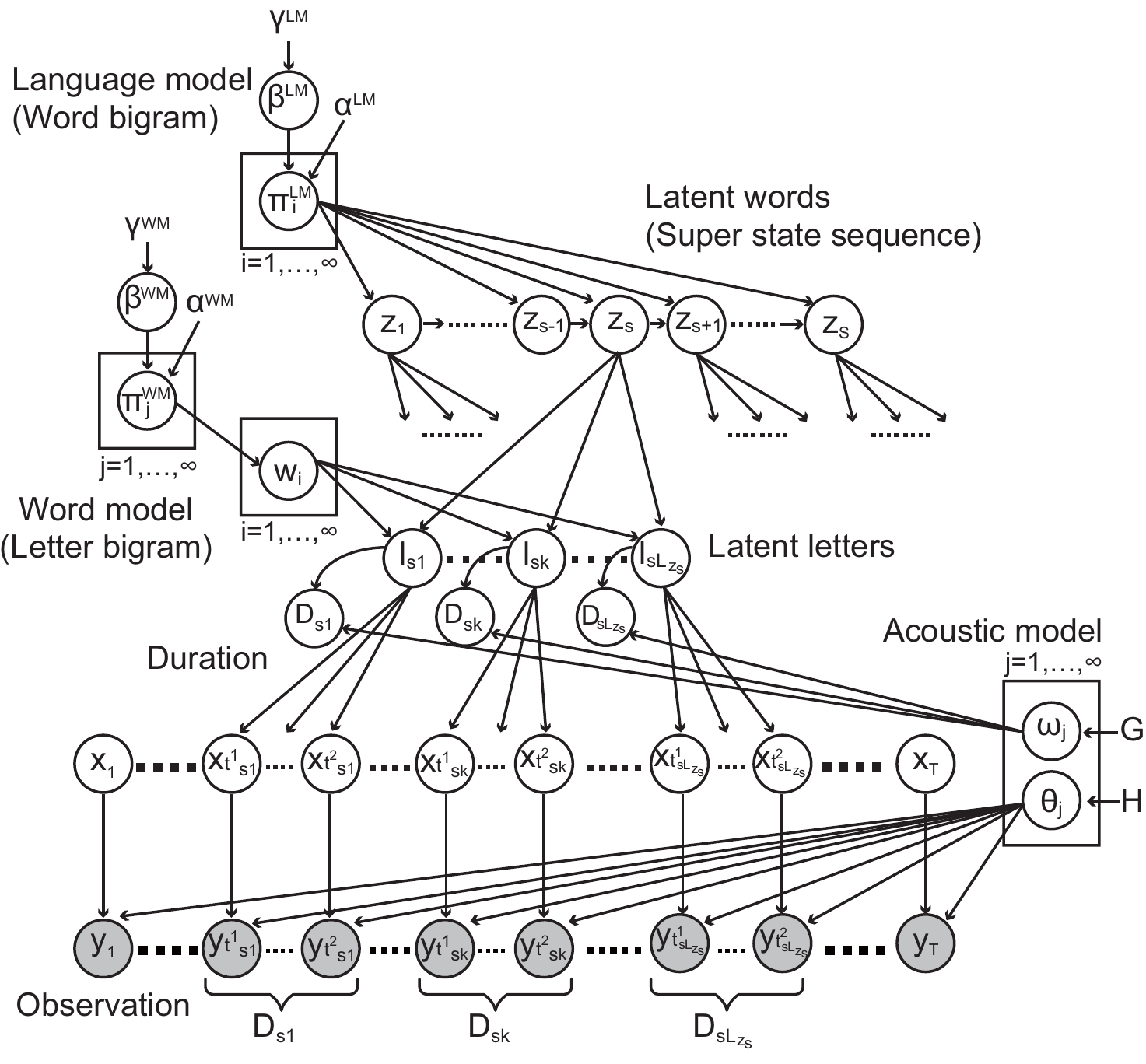}
  \caption{Graphical model of HDP-HLM \citep{Taniguchi2016}}
  \label{fig:graphical_model}
\end{figure*}
Whereas HDP-HMM assumes that the latent variable transits between them following Markov process, HDP-HLM assumes that the latent variable, index of phoneme, transits according to the word bigram language model. 
In HDP-HSMM, a superstate persists for a certain duration determined by the duration distribution and outputs observation using a corresponding emission distribution; meanwhile, in HDP-HLM, a latent word persists for a certain duration, and the model output observations with a sequential transition of latent letters, i.e., phonemes. Note that in the HDP-HLM terminology, the variable corresponding to a phoneme is called a latent letter; the variable corresponding to a word is called a latent word. 

As HMM-based ASR has language and acoustic models, HDP-HLM has both these as latent variables in its generative model. Because of the nature of Bayesian nonparametrics, i.e., Dirichlet process prior, HDP-HLM can determine the number of phonemes and words through the inference process. It is not necessary to fix the number of phonemes and words, i.e., the number of latent letters and words, beforehand. 

In the graphical model, the $s$-th latent word corresponds to superstate $z_s$. 
Superstate $z_s = i$ has a sequence of latent letters $w_{i} = (w_{i1}, \ldots, w_{ik}, \ldots, w_{iL_i})$; here, $w_{ik}$ is the index of the $k$-th latent letter of the $i$-th latent word. $L_i$ represents the string length of $w_i$.
The generative process of HDP-HLM is as follows; 
\begin{align}
  \beta^{LM} &\sim \GEM(\gamma^{LM})\\
  \pi^{LM}_i &\sim \DP(\alpha^{LM}, \beta^{LM}) && i = 1, 2, \dots, \infty\\
  \beta^{WM} &\sim \GEM(\gamma^{WM})\\
  \pi_j^{WM} &\sim \DP(\alpha^{WM}, \beta^{WM}) && j = 1, 2, \dots, \infty\\
  w_{ik} &\sim \pi_{w_{ik-1}}^{WM} && i = 1, 2, \dots, \infty &&& k = 1, 2, \dots, L_{i}\\
  (\theta_j, \omega_j) &\sim H \times G && j = 1, 2, \dots, \infty\\
  z_s &\sim \pi^{LM}_{z_{s-1}} && s = 1, 2, \dots, S\\
  l_{sk} &\sim w_{z_sk} && s = 1, 2, \dots, S
  &&& k = 1, 2, \dots, L_{z_s}
\end{align}
\begin{align}
  D_{sk} &\sim g(\omega_{l_{sk}}) && s = 1, 2, \dots, S && k = 1, 2, \dots, L_{z_s}\\
  x_t &= l_{sk} && t = t_{sk}^1 , \ldots , t_{sk}^2\nonumber\\
 &&& t_{sk}^1 = \sum_{s' < s} D_{s'} + \sum_{k' < k} D_{sk'} +1 &&t_{sk}^2 = t_{sk}^1 + D_{sk} - 1\\
  y_t &= h(\theta_{x_t}) && t = 1, 2,  \ldots , T
\end{align} 
Here, $\GEM$ represents a stick breaking process (SBP), and $\DP$ represents Dirichlet process (DP). 
Here, $\beta^{WM}$ represents the based measure of Dirichlet process for word model, $\alpha^{WM}$ and $\gamma^{WM}$ are hyperparameters of DP and  SBP.
A word model is a prior distribution of a sequence of latent letters composing a latent word. 
$\DP (\alpha^{WM}, \beta^{WM})$ generates a transition probability, $\pi^{WM}_j$ which is a categorical distribution over the subsequent latent letter of the $j$-th latent letter.
Similarly, $\beta^{LM}$, $\DP (\alpha^{LM}$, and $\beta^{LM})$ represent the based measure of Dirichlet process for language model and hyperparameters of DP and SBP, respectively.
$\DP (\alpha^{LM}, \beta^{LM})$ generates a transition probability $\pi^{LM}_i$; it is a categorical distribution over the subsequent latent letter of the $i$-th latent letter.
The notations $LM$ and $WM$ represent language and word models, respectively.
The emission distribution $h$ and duration distribution $g$ have parameters $\theta_j$ and  $\omega_j$ 
 drawn from the base measures $H$ and $G$, respectively.
The variable $z_s$ is the $s$-th word in the latent word sequence.
Moreover, $D_s$ is the duration of $z_s$, $l_{sk} = w_{z_sk}$ is the $k$-th latent letter of the $s$-th latent word, and $D_{sk}$ is its duration. 
Variables $y_t$ and $x_t$ represent the observation and latent state corresponding to a latent letter at time $t$.
The times $t_{sk}^1$ and $t_{sk}^2$ represent the start time and end time, respectively, of $l_{sk}$.

If we assume the duration distribution of a latent letter to follow a Poisson distribution, the model exhibits an effective mathematical feature because of the reproductive property of Poisson distributions.
The duration $D_{sk}$ is drawn from $g(\omega_{l_{sk}})$.
Therefore, the duration of $w_{z_s}$ is $D_s = \sum_{k = 1}^{L_{z_s}} D_{sk}$. 
If we assume $D_{sk}$ to follow Poisson distribution, i.e., $g$ is a Poisson distribution, $D_s$ also follows Poisson distribution. In this case, the parameter of the Poisson duration distribution of $w_{z_s}$ becomes $\sum_{k = 1}^{L_{z_s}} \omega_{l_{sk}}$.
The observation $y_t$ corresponding to $x_t = l_{s(t)k(t)}$ is generated from $h(\theta_{x_t})$; here, $s(t)$ and $k(t)$ are mappings that indicate the corresponding word $s$ and letter $k$ at time $t$.

Following the process described above, HDP-HLM can generate time series data exhibiting a latent double articulation structure.
In this study, we assumed that the observation $y_t$ corresponds to the acoustic features.
In summary, $\{ \omega_j, \theta_j \}_{j=1,2, \ldots , \infty}$ represents acoustic models, and $\{ \pi^{LM}_i , w_i \}_{i=1,2, \ldots , \infty}$ represents language models. The inference of the latent variables of this generative model corresponds to the simultaneous discovery of phonemes and words.

An inference procedure for HDP-HLM was proposed in~\citep{Taniguchi2016}. 
This procedure is based on the blocked Gibbs sampler for HDP-HSMM proposed by Johnson~\citep{Johnson2013}.
The pseudo code of the procedure is described in Algorithm~\ref{alg:bgibbs}. 
In this paper, we omit the details of the procedure. For further details, please refer to the original paper~\citep{Taniguchi2016}.

\begin{algorithm}[H]
\caption{Blocked Gibbs sampler for HDP-HLM}
\label{alg:bgibbs}
\begin{algorithmic}
\STATE Initialize all parameters.
\STATE Observe $M$ time series data $\{y^m_{1:T_m}\}_{m\in\{1, 2, \ldots, M\}}$. 
\REPEAT
\FOR{$m=1$ to $M$}
\STATE // Backward filtering procedure
\STATE For each ${i\in\{1, 2, \ldots, N\}}$, initialize messages $B_T(i)=1$.
\FOR{$t=T$ to $1$} 
\STATE For each ${i\in\{1, 2, \ldots, N\}}$, compute backward messages $B_{t-1}(i)$ and $B_{t-1}^*(i)$ (see \cite{Taniguchi2016})
\ENDFOR
\STATE // Forward sampling procedure
\STATE Initialize $s=1$ and $D^{\text{sum}}_s=0$
\WHILE{$D^{\text{sum}_s} < T_m$} 
\STATE // Sampling a superstate representing a latent word
\STATE $z_s \sim p(z_s \mid y^m_{1:T_m}, z_{s-1}, F_{D^{\text{sum}}_s}=1 )$ 
\STATE // Sampling duration of the superstate  
\STATE $D_s \sim p(D_s | z_s , F_{D^{\text{sum}}_s}=1  )$ 
\STATE  $D^{\text{sum}}_{s+1} \leftarrow D^{\text{sum}}_s + D_s$
\STATE $s \leftarrow s+1$
\ENDWHILE
\STATE $S^m \leftarrow s-1$
\STATE // Sampling a tentative latent letter sequence 
\FOR{$s=1$ to $S^m$} 
\STATE $ \bar{w}^m_s \sim P(w | y^m_{D^{\text{sum}}_{s-1}+1:D^{\text{sum}}_s}, \{\pi_j^{WM}, \omega_j, \theta_j \}_{j=1,2,\ldots,J})$ 
\ENDFOR
\ENDFOR
\STATE // Update model parameters
\STATE Sample acoustic model parameters $\{ \omega_j, \theta_j\}$ on the basis of tentatively sampled latent letter sequences $\{ \bar{w}^m_s\}$.
\STATE Sample language model parameter $\{\pi_i^{LM}\}, \beta^{LM}$ on the basis of sampled super states , i.e., latent words. 
\STATE Sample a word inventory $\{w_i\}_{i = 1, 2, \ldots, N}$ using sampling importance re-sampling procedure (see \cite{Taniguchi2016}).
\STATE Sample a word model $\{\pi_i^{WM}\}, \beta^{WM}$ on the basis of the sampled word inventory  $\{w_i\}_{i = 1, 2, \ldots, N}$. 
\UNTIL{a predetermined exit condition is satisfied.}
\end{algorithmic}
\end{algorithm}

\subsection{Deep sparse auto-encoder with parametric bias}
\subsubsection{Deep sparse auto-encoder}
In the previous paper~\citep{Taniguchi2016b}, features extracted using DSAE were used as the input of NPB-DAA.
DSAE is a representation learning method. It consists of several sparse autoencoders (SAEs)~\citep{ng2011sparse}.
By stacking several autoencoders and assigning penalty terms to the loss function for improving robustness and sparsity, DSAE is obtained. 
In DSAE, each SAE attempts to minimize the reconstruction errors and learn efficient and essential representations of the input data (speech signals in this study).

\begin{figure*}[t]
  \centering
  \includegraphics[width = 0.7\linewidth]{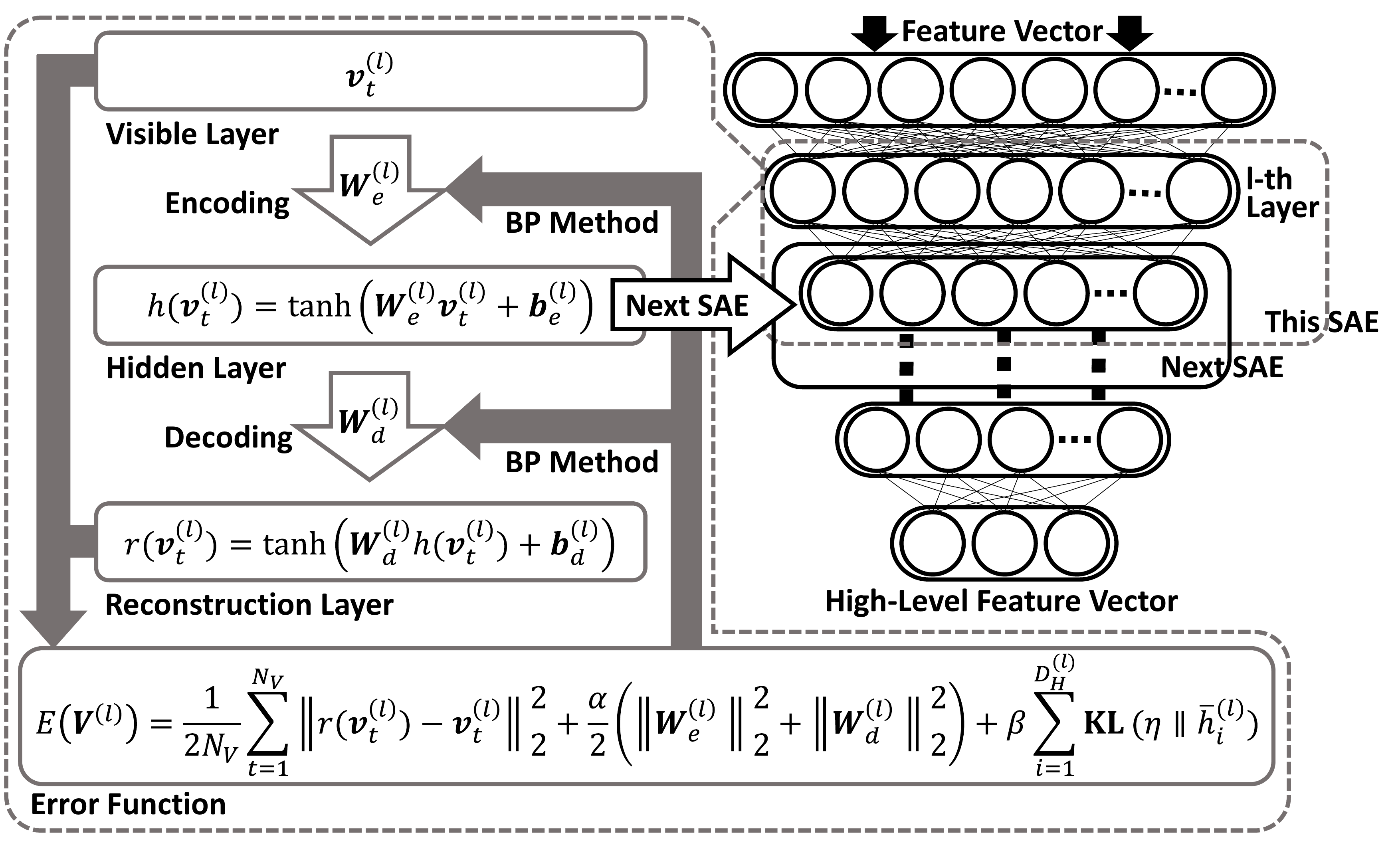}
  \caption{Overview of DSAE}
  \label{fig:dsae}
\end{figure*}
Figure~\ref{fig:dsae} shows an overview of DSAE.
In this study, we assumed that the original input of speech signals are converted into Mel frequency cepstral coefficients (MFCC) following the process described in the previous work~\citep{Taniguchi2016b}.
The time series data is obtained as a matrix ${\bf O}\in\mathbb{R}^{D_{O}\times N_{O}}$; here, $N_O$ is the number of data.
The acoustic feature at time $t$ is represented by ${\bf o}_t\in\mathbb{R}^{D_{O}}$ as follows:
\begin{eqnarray}
 {\bf o}_t=(o_{t,1} ,o_{t,2} ,\dots , o_{t,D_O})^T
 \end{eqnarray}
 where $D_O$ represents the dimension of vector ${\bf  o}_t$. 
 
In this study, the hyperbolic tangent function ${\bf tanh}(\cdot)$  was used as the activation function of SAE.  
To fit the input data to the range of ${\bf tanh}(\cdot)$ for reconstruction, the input vector ${\bf  o}_t$ was normalized as follows:
\begin{eqnarray}
{\bf v}_t=(v_{t,1} ,v_{t,2} ,\dots , v_{t,D_O})^T && v_{t,d}=2\Big(\frac{o_{t,d}-O_{\small\mbox{min},d}}{O_{\small\mbox{max},d}-O_{\small\mbox{min},d}}\Big)-1
 \end{eqnarray}
 where $O_{\small\mbox{max},d}$ and $O_{\small\mbox{min},d}$ are the maximum and minimum values, respectively, of the $d$-th dimension of
 all the data $\bf o \in \bf O$ .

Each SAE has an encoder and a decoder. The encoder of the $l$-th SAE in DSAE is 
\begin{eqnarray}
{\bf h}^{(l)}_t={\bf tanh}({\bf W}_e^{(l)}{\bf v}^{(l)}_t+{\bf b}_{e}^{(l)})\label{eq:10}.
\end{eqnarray}
Following this function, about the $t$-th data, a vector of the $l$-th layer  ${\bf v}^{(l)}_t$ is transformed to a vector of the $l$-th hidden layer ${\bf h}^{(l)}_t\in\mathbb{R}^{D_H^{(l)}}$.
Each decoder is represented as follows: The vector of the $l$-th layer ${\bf r}^{(l)}_t\in\mathbb{R}^{D_V^{(l)}}$ is obtained from the vector of the $l$-th reconstruction layer.
\begin{eqnarray}
{\bf r}^{(l)}_t={\bf tanh}({\bf W}_{d}^{(l)}{\bf h}({\bf v}^{(l)}_t)+{\bf b}_{d}^{(l)})\label{eq:11}
\end{eqnarray}
where  ${\bf W}_e^{(l)}\in\mathbb{R}^{D_H^{(l)}\times D_V^{(l)}}$ in  (\ref{eq:10}) is the weight matrix and ${\bf b}_{e}^{(l)}\in\mathbb{R}^{D_H^{(l)}}$ is the bias of the encoder. 
Moreover, $\mathbb{R}^{D_V^{(l)}}$ and $\mathbb{R}^{D_H^{(l)}}$ represent the dimensions of the input and hidden layers, respectively.
Similarly,${\bf W}_d^{(l)}\in\mathbb{R}^{D_V^{(l)}\times D_H^{(l)}}$ in (\ref{eq:11} ) is the weight matrix of the decoder, and ${\bf b}_{d}^{(l)}\in\mathbb{R}^{D_V^{(l)}}$ is the bias.

The loss function was defined as follows: 
\begin{eqnarray}
E({\bf V}^{(l)})=\frac{1}{2N_V}\sum_{t=1}^{N_V}||{\bf r}^{(l)}_t-{\bf v}^{(l)}_t||_{2}^{2}+\frac{\alpha}{2}(||{\bf W}_{e}^{(l)}||_{2}^{2}+||{\bf W}_{d}^{(l)}||_{2}^{2})+\beta\sum^{D_{H}^{(l)}}_{i=1}\mbox{KL}(\eta||{\bar{h}}_i^{(l)})
\label{eq:12}
\end{eqnarray}
Because the dimensions of the weight matrices ${\bf W}_{e}^{(l)}$ and ${\bf W}_{d}^{(l)}$ are high, it was necessary to prevent the penalty terms ${\bf W}_{e}^{(l)}$, ${\bf W}_{d}^{(l)}$ (L2 norm) and $\beta\sum^{D_{H}^{(l)}}_{i=1}\mbox{KL}(\eta||{\bar{h}}_i^{(l)})$ (sparse term); this is the Kullback--Leibler divergence between the two Bernoulli distributions having $\eta$ and ${\bar{h}}_i^{(l)}$ as their parameters. This type of DSAE is introduced in~\citep{ng2011sparse}. The following are details of the sparse term:  
\begin{eqnarray}
\mbox{KL}(\eta||{\bar{h}}_i^{(l)})=\eta\log\frac{\eta}{{\bar{h}}_i^{(l)}}+(1-\eta)\log\frac{1-\eta}{1-{\bar{h}}_i^{(l)}}&&\boldmath{{\bar h}}^{(l)}_i=\frac{1}{2}\Big(1+\frac{1}{N_V}\sum_{t=1}^{N_V}h^{(l)}_{t,i}\Big)
\label{eq:13}
\end{eqnarray}
where $\eta\in\mathbb{R}$ is a parameter that regulates sparsity. Moreover, ${\bar{h}}_i^{(l)}$ represents the average of the $i$-th dimension's activation. The vector $\boldmath{ {\bar{\bf h}}}^{(l)}\in\mathbb{R}^{D_H^{(l)}}$ ${\bar{h}}_i^{(l)}$ is defined by combining ${\bar{h}}_i^{(l)}$. In this study, to calculate the sparse term,
$\boldmath{{\bar{\bf h}}}^{(l)}$ was normalized from $(-1,1)$ to $(0,1)$; this was because ${\bf tanh}(\cdot)$ was used as an activation function.
To optimize the DSAE, simply back-propagation method was used~\citep{rumelhart1985learning}. 

As described above, we could obtain the weight matrices ${\bf H}^{(l)}=\big({\bf h}^{(l)}_1,\dots,{\bf h}^{(l)}_t\big)\in\mathbb{R}^{D_{H}^{(l)}\times N_V}$ for obtaining ${\bf V}^{(l+1)}\in\mathbb{R}^{D_{H}^{(l)}\times N_V}$. By stacking the optimized SAE's, high-level feature representations could be obtained.

\subsubsection{DSAE-PBHL}

\begin{figure}[t]
  \centering
    \includegraphics[width=0.9\linewidth]{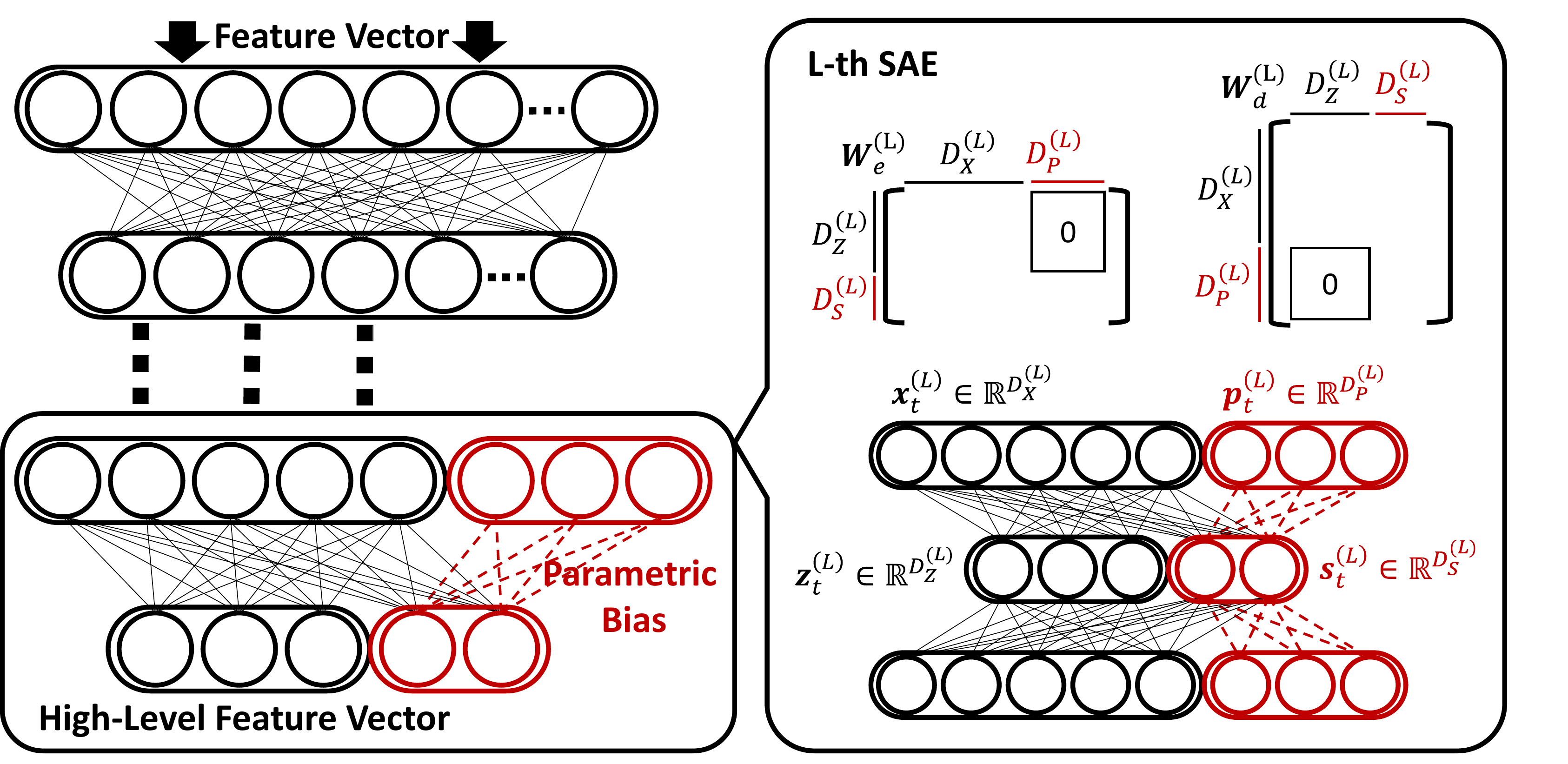}
  \caption{Overview DSAE-PBHL}
  \label{fig:dp}
\end{figure}

This section describes DSAE-PBHL that is aimed to subtract speaker-dependent features in the latent space. 

DSAE-PBHL is a DSAE that has a final layer; a part of this layer receives speaker index information from the other network.
The layer is used to subtract speaker-dependent information in a self-organizing manner.
Figure~\ref{fig:dp} shows an overview of DSAE-PBHL. 
The $L$-th layer, i.e., the final layer, receives parametric bias input from a different network (see the right nodes of the network in Figure~\ref{fig:dp}).  

However, the vital aspect of DSAE-PBHL is that a part of nodes in the final layer receives a projection from the network representing speaker index information.

The input vector ${\bf v}_t^{(L)}\in\mathbb{R}^{D_{V}^{(L)}}$ consists of the parametric bias ${\bf p}_t^{(L)}\in\mathbb{R}^{D_{P}^{(L)}}$ and a vector ${\bf x}_t^{(L)}\in\mathbb{R}^{D_{X}^{(L)}}$ obtained from the $(L-1)$-th SAE.
\begin{eqnarray}
{\bf v}_t^{(L)}=({\bf x}_t^{(L)}, {\bf p}_t^{(L)})^T\in\mathbb{R}^{D_{V}^{(L)}}
\end{eqnarray}
where $D_{X}^{(L)}$ and $D_{P}^{(L)}$ represents the dimensions of ${\bf x}_t^{(L)}$ and ${\bf p}_t^{(L)}$, respectively. Note that $D_{V}^{(L)}=D_{X}^{(L)}+D_{P}^{(L)}$. 

Next, the vector of the $L$-th hidden layer ${\bf h}_t^{(L)}\in\mathbb{R}^{D_{H}^{(L)}}$, ${\bf x}_t^{(L)}$,${\bf p}_t^{(L)}$ is defined using ${\bf z}_t^{(L)}\in\mathbb{R}^{D_{Z}^{(L)}}$,${\bf s}_t^{(L)}\in\mathbb{R}^{D_{S}^{(L)}}$ as follows:
\begin{eqnarray}
{\bf h}_t^{(L)}=({\bf z}_t^{(L)}, {\bf s}_t^{(L)})^T\in\mathbb{R}^{D_{H}^{(L)}}
\end{eqnarray}
where $D_{Z}^{(L)}$ and $D_{S}^{(L)}$ represent the dimensions of ${\bf z}_t^{(L)}$ and  ${\bf s}_t^{(L)}$, respectively. Note that $D_{H}^{(L)}=D_{Z}^{(L)}+D_{S}^{(L)}$.

The encoder of the $L$-th SAE used (\ref{eq:10}) similarly as the general DSAE. However, the weight matrix of the encoder was trained to map the input vectors ${\bf x}_t^{(L)}$ and ${\bf p}_t^{(L)}$ to the latent vectors ${\bf z}_t^{(L)}$ and ${\bf s}_t^{(L)}$ in the hidden layer and generate speaker-independent feature representation and speaker-identifiable representation. 
\begin{eqnarray}
{\bf W}_{e}^{(L)}=
\begin{pmatrix}
{\bf W}_{\bf{z,x}}^{(L)} & {\bf W}_{\bf{z,p}}^{(L)} \\
{\bf W}_{\bf{s,x}}^{(L)} & {\bf W}_{\bf{s,p}}^{(L)} \\
\end{pmatrix}
\in\mathbb{R}^{D_H^{(L)}\times D_V^{(L)}}
\end{eqnarray}
where,${\bf W}_{\bf{z,x}}^{(L)}\in\mathbb{R}^{D_Z^{(L)}\times D_X^{(L)}}$,${\bf W}_{\bf{z,p}}^{(L)}\in\mathbb{R}^{D_Z^{(L)}\times D_P^{(L)}}$,${\bf W}_{\bf{s,x}}^{(L)}\in\mathbb{R}^{D_S^{(L)}\times D_X^{(L)}}$,${\bf W}_{\bf{s,p}}^{(L)}\in\mathbb{R}^{D_S^{(L)}\times D_P^{(L)}}$, ${\bf W}_{\bf{z,p}}^{(L)}=\text{{\Large{0}}}$.

Similarly, the decoder function (\ref{eq:11}) was used, and the weight matrix of the decoder function was modified as follows:
\begin{eqnarray}
{\bf W}_{d}^{(L)}=
\begin{pmatrix}
{\bf W}_{\bf{x,z}}^{(L)} & {\bf W}_{\bf{x,s}}^{(L)} \\
{\bf W}_{\bf{p,z}}^{(L)} & {\bf W}_{\bf{p,s}}^{(L)} \\
\end{pmatrix}
\in\mathbb{R}^{D_V^{(L)}\times D_H^{(L)}}
\end{eqnarray}
where, ${\bf W}_{\bf{x,z}}^{(L)}\in\mathbb{R}^{D_X^{(L)}\times D_Z^{(L)}}$,${\bf W}_{\bf{x,s}}^{(L)}\in\mathbb{R}^{D_X^{(L)}\times D_S^{(L)}}$,${\bf W}_{\bf{p,z}}^{(L)}\in\mathbb{R}^{D_P^{(L)}\times D_Z^{(L)}}$,${\bf W}_{\bf{p,s}}^{(L)}\in\mathbb{R}^{D_P^{(L)}\times D_S^{(L)}}$ , and ${\bf W}_{\bf{p,z}}^{(L)}=\text{{\Large{0}}}$.

Furthermore, the error function and optimization method were identical to those in the general DSAE. 

After the training phase, ${\bf z}_t^{(L)}$ was obtained by excluding  ${\bf s}_t^{(L)}$ from the vector of the $L$-th hidden layer, ${\bf h}_t^{(L)}$ and was used as a feature vector, i.e., observation, of NPB-DAA.

The reason we considered it likely that ${\bf z}_t^{(L)}$ encoded speaker-independent feature representation is that the network was trained to cause  ${\bf s}_t^{(L)}$ to have speaker-identifiable representation; this was because ${\bf s}_t^{(L)}$ alone had to contribute to reconstructing the speaker index information, i.e., parametric bias.
As Figure~\ref{fig:dp} shows, ${\bf s}_t^{(L)}$ was connected only to the input of parametric bias, i.e., speaker index.
If ${\bf z}_t^{(L)}$ involves speaker-dependent information that can be used to predict the speaker index, the representation is redundant. Therefore, such speaker-dependent information is likely to be mapped onto ${\bf s}_t^{(L)}$. As a result, it is likely that ${\bf z}_t^{(L)}$ becomes encoding information that does not contribute to the speaker identification task (i.e., it becomes speaker-independent information).

\section{Experiment}\label{sec:3}
To evaluate the proposed method, we conducted two experiments. First, we tested whether DSAE-PBHL can extract speaker-independent feature representations using speech signals representing isolated Japanese vowels and an elementary clustering method. Secondly, we tested whether NPB-DAA with DSAE-PBHL can successfully perform unsupervised phoneme and word discovery from speech signals obtained from multiple speakers.
\subsection{Common conditions}
In the following two experiments, we used the common dataset. The procedure of creating the data is identical to that in the previous related papers~\citep{Taniguchi2016,Taniguchi2016b}. 

We asked two male and two female Japanese speakers to read 30 artificial sentences aloud once at a natural speed and recorded their voice using a microphone. Totally, 120 audio data items were recorded. 
We name the two female datasets as K-DATA and M-DATA and the two male datasets as H-DATA and N-DATA, respectively.

The 30 artificial sentences were prepared using five artificial words \{aioi, aue, ao, ie, uo\} consisting of five Japanese vowels \{a, i, u, e, o\}. By reordering the words, 25 two-word sentences, e.g., ``ao aioi,'' ``uo aue,'' and ``aioi aioi,'' and five three-word sentences, i.e., ``uo aue ie,'' ``ie ie uo,'' ``aue ao ie,'' ``ao ie ao,'' and ``aioi uo ie,'' were prepared. The set of two-word sentences comprised of all the feasible pairs of the five words ($5 \times 5 =25$). The set of three-word sentences were determined manually.

The input speech signals were provided as MFCCs, which have been widely used in ASR studies.
The recorded data were encoded into $39$-dimensional MFCC time series data using the HMM Toolkit (HTK).\footnote{Hidden Markov Model Toolkit: http://htk.eng.cam.ac.uk/} The frame size and shift were set to $25$ ms and $10$ ms, respectively. Twelve-dimensional MFCC data were obtained as the input data by eliminating the power information from the original 13-dimensional MFCC data. As a result, 12-dimensional time series data at a frame rate of $100$ Hz were obtained.

In DSAE-PBHL, 39-dimensional MFCC was compressed by DSAE, whose variation in the dimensions was $39 \rightarrow 20 \rightarrow 10 \rightarrow 6.$ 
The speaker index was provided to the final layer as a four-dimensional input. 
In the final layer, the dimensions of ${\bf z}_t^{(L)}$ and ${\bf s}_t^{(L)}$ were three and three, respectively. We used ${\bf z}^{(L)}$ as an input of clustering methods, e.g., k-means, GMM, and NPB-DAA.

In DSAE, the 39-dimenssional MFCC was compressed by DSAE, whose variation in the dimensions was $39 \rightarrow 20 \rightarrow 10 \rightarrow 6 \rightarrow 3.$ 
The parameters in DSAE were set as $\alpha = 0.003$, $\beta = 0.7$, and $\eta = 0.5$. 

\subsection{Experiment 1: Vowel clustering based on DSAE-PBHL}
This experiment evaluates if the DSAE-PBHL can extract speaker-independent representations from the perspective of a phoneme clustering task rather than a word discovery task.
\subsubsection{Conditions}
For quantitative evaluation, we applied two elementary clustering methods (k-means and GMM) to the extracted feature vectors to examine whether the DSAE-PBHL extracts speaker-independent feature representations.
If the elementary clustering methods can identify clusters corresponding to each vowel, it implies that each phoneme forms clustered distributions to a certain extent. The clustering performance was quantified with the adjusted Rand index (ARI), which is a standard evaluation criterion of clustering.

We also tested three types of coding of parametric bias, i.e., sparse coding and codings 1 and 2 (Table~\ref{tbl:exp_ari2}).

As a baseline method, we employed DSAE.

Furthermore, we applied DSAE and the clustering methods separately to the four datasets (H-DATA, K-DATA, M-DATA, and N-DATA) and calculated the average of the ARI. This result can be considered as an upper limit of the performance.

The codes of scikit-learn\footnote{http://scikit-learn.org/stable/} were used for k-means and GMM. 
The number of clusters of the methods was fixed as five, i.e., the exact number. With regard to the other hyperparameters, the default settings of scikit-learn was used.

\subsubsection{Results}
Table~\ref{tbl:exp_ari2} presents the ARI averaged over 20 trials for k-means and GMM and for each method.
This result demonstrates that DSAE-PBHL exhibited significantly higher performance than DSAE and MFCC in the representation learning of acoustic features from multiple speakers, in phoneme clustering. 
Among the three coding methods, sparse coding, i.e., one-hot vector, achieved the bests core.
In numerous cases in deep learning, sparse coding exhibits effective characteristics. Therefore, this result appears consistent. However, even in different cases of encoding methods, DSAE-PBHL outperformed other methods.
As was considered likely, DSAE-PBHL did not attain the upper limit, although, it reduced the difference.

\begin{table}[bt]
\caption{ARI in phoneme clustering task}
\label{tbl:exp_ari2}
\begin{center}
\begin{tabular}{|c||c|c||c|}
\hline
Method & \shortstack{k-means} & \shortstack{GMM} & PB: [H-PB], [K-PB], [M-PB], [N-PB]\\
\hline\hline
DSAE-PBHL (Sparse Coding)& \underbar{\bf 0.536} & \underbar{\bf 0.519} & [0,0,0,1], [0,0,1,0], [0,1,0,0], [1,0,0,0]\\
\hline
DSAE-PBHL (Coding 1)& 0.514 & 0.429 & [0,0,0,1], [0,0,1,0], [0,0,1,1], [0,1,0,0]\\
\hline
DSAE-PBHL (Coding 2)& 0.448 & 0.362 & [0,0,1,1], [0,1,1,0], [1,1,0,0], [1,0,0,1]\\
\hline\hline
DSAE & 0.212 & 0.222 &\\
\hline
MFCC & 0.243 & 0.182 &\\
\hline\hline
Upper Limit & \bf0.626 & \bf0.599 &\\
\hline
\end{tabular}
\end{center}
\end{table}

\begin{figure}[t]
  \centering
    \includegraphics[width=\linewidth]{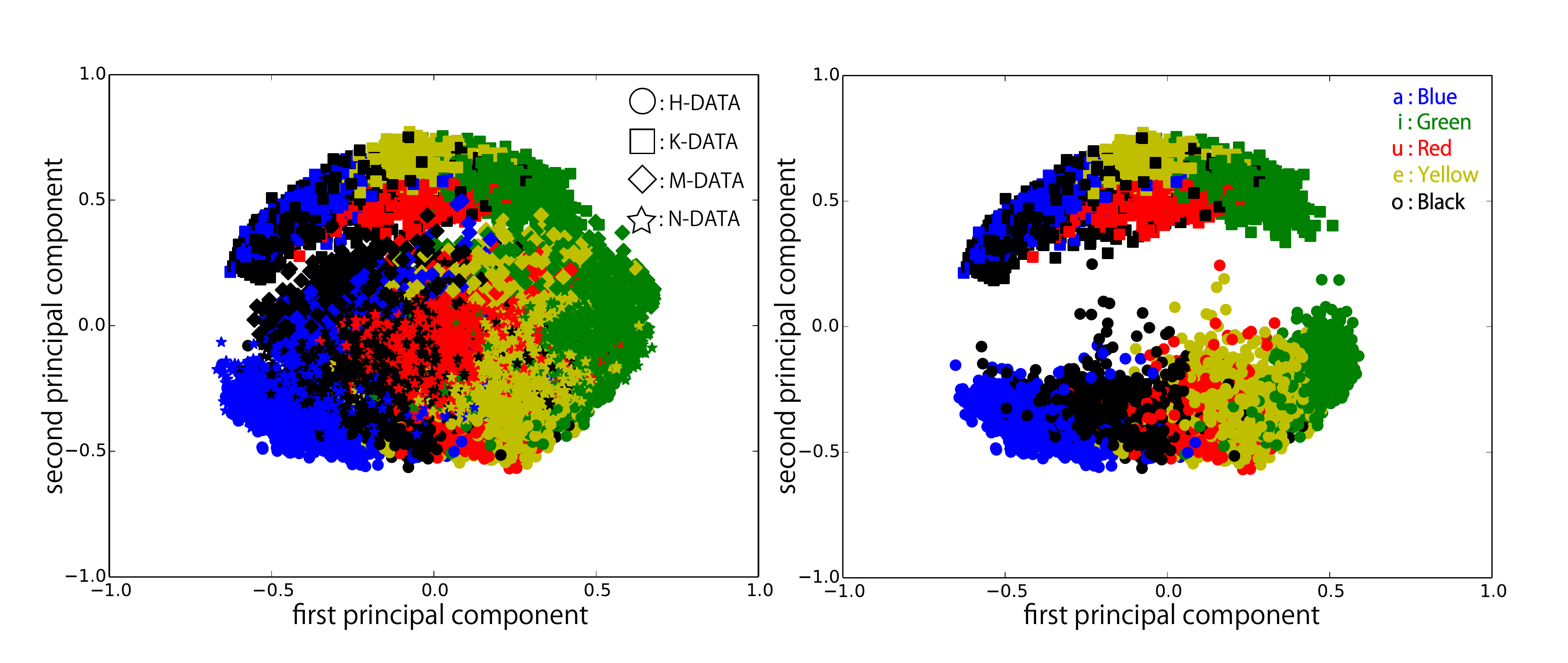}
  \caption{Feature representations extracted by DSAE visualized using PCA. (Left) all data, (right) H-DATA and K-DATA.}
  \label{fig:d}
\end{figure}

\begin{figure}[t]
  \centering
    \includegraphics[width=\linewidth]{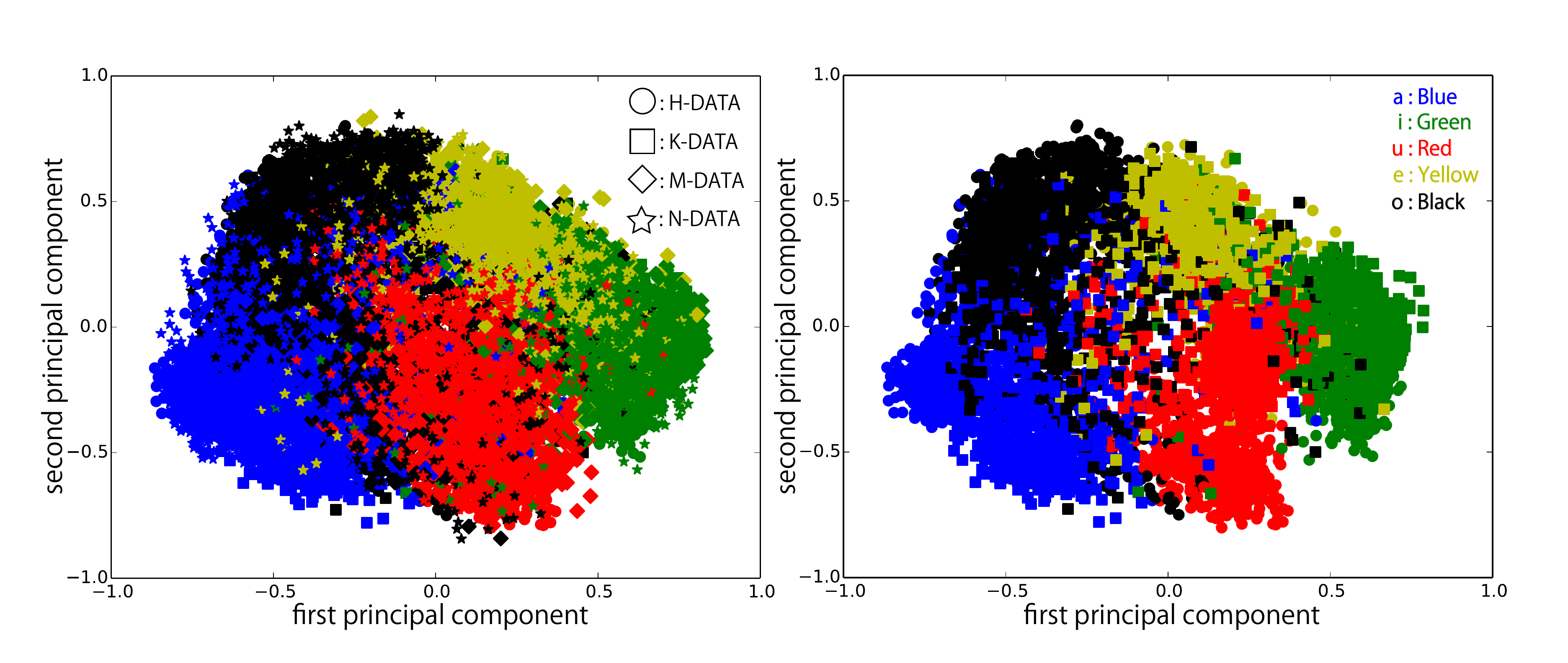}
  \caption{Feature representations extracted by DSAE-PBHL (Sparse Coding) visualized using PCA. (Left) all data, (right) H-DATA and K-DATA.}
  \label{fig:ex1}
\end{figure}

\begin{figure}[t]
  \centering
    \includegraphics[width=\linewidth]{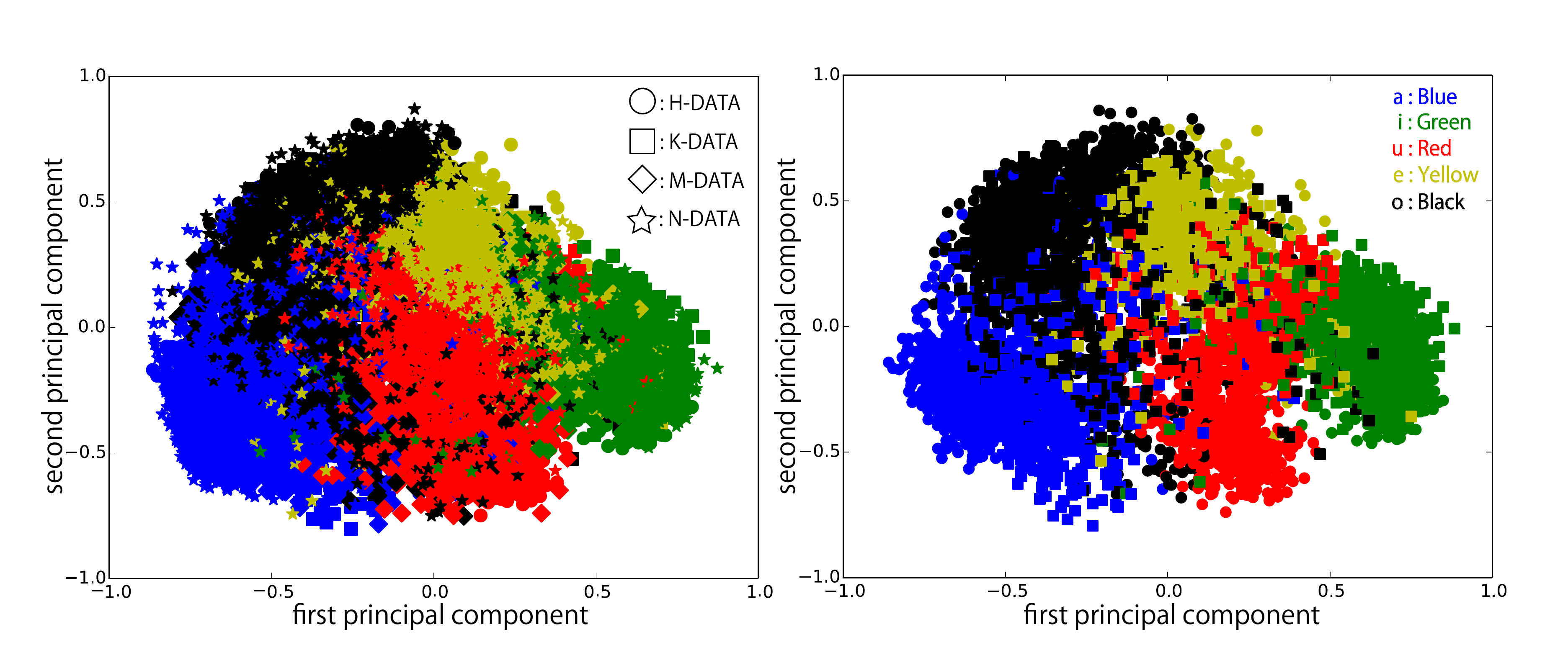}
  \caption{Feature representations extracted by DSAE-PBHL (Coding 1) visualized with PCA. (Left) all data, (right) H-DATA and K-DATA.}
  \label{fig:ex2}
\end{figure}

\begin{figure}[t]
  \centering
    \includegraphics[width=\linewidth]{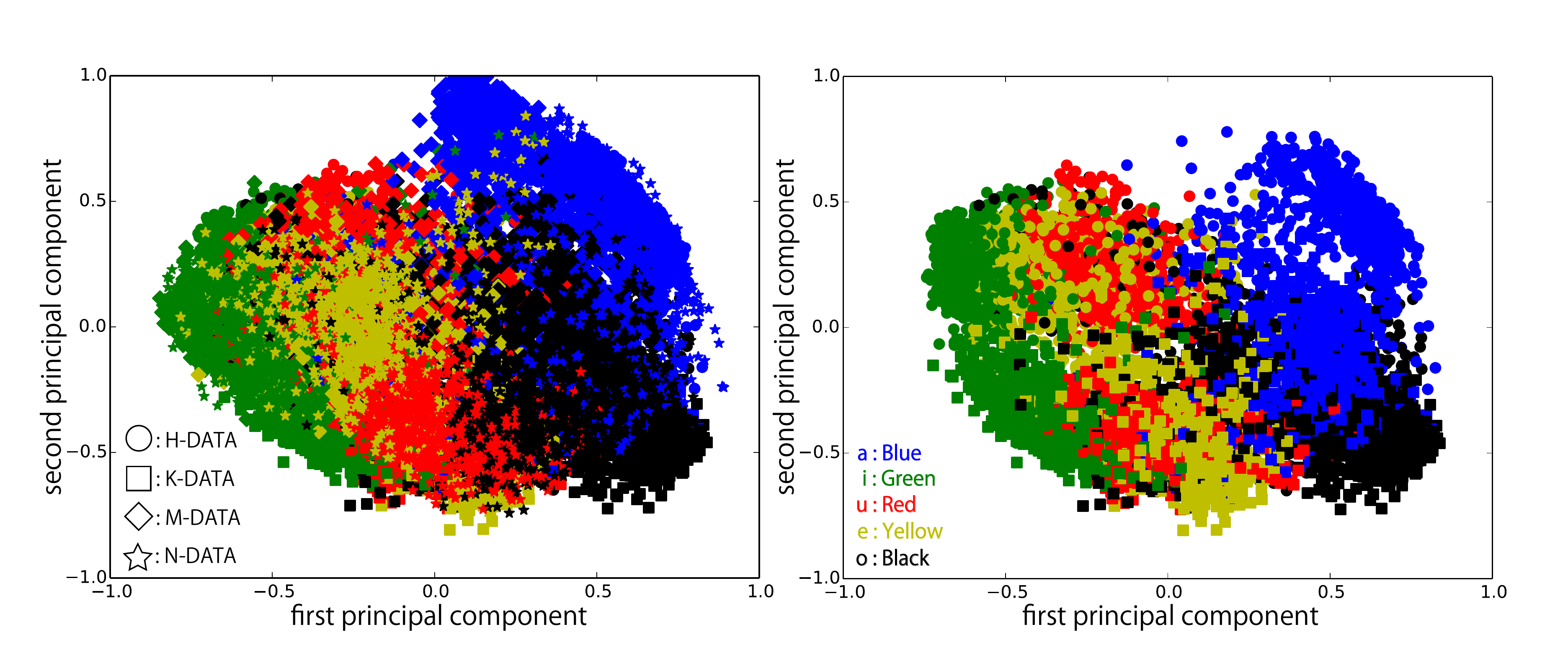}
  \caption{Feature representations extracted by DSAE-PBHL (Coding 2) visualized with PCA. (Left) all data, (right) H-DATA and K-DATA.}
  \label{fig:ex3}
\end{figure}

Figures \ref{fig:d}, \ref{fig:ex1}, \ref{fig:ex2} and \ref{fig:ex3} visualizes feature representations extracted by DSAE and DSAE-PBHL with three types of codings.
The final three-dimensional representation is mapped to two-dimensional space using principal component analysis (PCA) for the purpose of visualization. 
In each figure, on the left is the scatter plot of the data from the four speakers, and the one on right is the scatter plot of the data from H-DATA and K-DATA, i.e., a male and a female speaker.

On the one hand, it was observed that DSAE formed speaker-dependent distributions (see Figure~\ref{fig:d}). 
For example, ``a'' from H-DATA and ``a'' from K-DATA formed entirely different clusters in the feature space.

On the other hand, DSAE-PBHL could form speaker-independent representation to a certain extent. 

\subsection{Experiment 2: simultaneous phoneme and word discovery from multiple speakers using NPB-DAA with DSAE-PBHL}
This experiment evaluates whether NPB-DAA with DSAE-PBHL can discover phonemes and words from speech signals from multiple speakers.

\subsubsection{Conditions}

The hyperparameters for the latent language model were set to $\gamma^{LM} = 10.0$ and $\alpha^{LM} = 10.0$; the maximum number of words was set to seven for weak-limit approximation. 
The hyperparameters of the duration distributions were set to $\alpha = 200$ and $\beta = 10$; those of the emission distributions were set to $\mu_0 = 0, \sigma^2_0 = 1.0, \kappa_0 = 0.01,$ and  $\nu_0 = 17 = ($dimension$ + 5)$.

The Gibbs sampling procedure was iterated 100 times for NPB-DAA. Twenty trials were performed using different random number seeds. 
Sparse coding of parametric bias was employed as the coding method of speaker index. 

We comepared NPB-DAA with DSAE-PBHL, NPB-DAA with MFCC, and NPB-DAA with DSAE.  
Similary as in Experiment 1, we calculated the performance of NPB-DAA with DSAE, which learns speakers separately, as an upper limit of the model.
Moreover, we used the off-the-shelf speech recognition system Julius\footnote{Julius: http://julius.sourceforge.jp/} having 
a pre-existing true dictionary consisting of \{aioi, aue, ao, ie, uo\} to output reference value of ARIs.
We used two types of Julius: one is the HMM-based model Julius, and the other is the deep neural network(DNN)-based Julius, namely Julius DNN.

\subsubsection{Resuluts}
Similarly as in Experiment 1, Table~\ref{tbl:exp_ari3} presents the ARIs for each condition.
The rows with (MAP) list the score when NPB-DAA exhibits the highest likelihood; the other rows list the average score of 20 trials.
The column SS represents the single speaker setting. Speech signals from different speakers are input separately and learned independently. This condition is considered as an upper limit of the proposed model.
The columns AM and LM illustrate whether the method uses pre-trained acoustic and language models, i.e., uses transcribed data, respectively.

This demonstrates that NPB-DAA with DSAE-PBHL (MAP), i.e., our proposed method, outperformed the previous models; however, it did not outperform the upper-limit method and Julius DNN.
However, it is noteworthy that NPB-DAA with DSAE outperformed Julius, which was trained in a supervised manner. 
\begin{table}[bt]
\caption{ARIs in phoneme and word discovery task}
\label{tbl:exp_ari3}
\begin{center}
\begin{tabular}{|c||c|c||c|c|c|}
\hline
Method & \shortstack{Letter ARI} & \shortstack{Word ARI} &SS &AM & LM \\
\hline\hline
NPB-DAA with DSAE-PBHL (MAP)& \underbar{\bf0.597} & \underbar{\bf0.373}&&&\\
\hline
NPB-DAA with DSAE-PBHL & 0.445 & 0.308&&&\\
\hline
NPB-DAA with DSAE (MAP)& 0.161 & 0.073&&&\\
\hline
NPB-DAA with DSAE & 0.234 & 0.139&&&\\
\hline
NPB-DAA with MFCC (MAP)& 0.281 & 0.115&&&\\
\hline
NPB-DAA with MFCC & 0.297 & 0.104&&&\\
\hline\hline
\shortstack{Upper Limit (speaker-dependence): \\NPB-DAA with DSAE (MAP)}& \bf0.621 & \bf0.627& \checkmark &&\\
\hline
\shortstack{Upper Limit (speaker-dependence) : \\NPB-DAA with DSAE} & 0.523 & 0.448& \checkmark &&\\
\hline\hline
\shortstack{Julius (triphone + word dictionary)} & 0.552 &  0.599& -- & \checkmark & \checkmark\\
\hline
\shortstack{Julius DNN (triphone + word dictionary)} & {\bf 0.693} & {\bf 0.791}& --  & \checkmark & \checkmark\\
\hline
\end{tabular}
\end{center}
\end{table}

This result indicates that DSAE-PBHL can reduce the adverse effect of obtaining speech signals from multiple speakers and that the simultaneous use of NPB-DAA can achieve direct phenome and word discovery from speech signals obtained from multiple speakers, to a certain extent.

\section{Conclusion}\label{sec:4}
This paper proposed a new method, NPB-DAA with DSAE-PBHL, for direct phoneme and word discovery from multiple speakers. In particular, DSAE-PBHL was developed to reduce the negative effect of speaker-dependent acoustic features in an unsupervised manner by using speaker index that is required to be obtained through another speaker recognition method.
This can be regarded as a more natural computational model of phoneme and word discovery by a human infant because it does not use transcription.
Human infants can acquire knowledge or phonemes and words from interactions his/her mother but as well as with other individuals surrounding him/her. We assumed that an infant can recognize and distinguish speakers by considering certain other features, e.g., visual face recognition. 
The study was aimed at enabling DSAE-PBHL to subtract speaker-dependent acoustic features and extract speaker-independent features. 
The first experiment demonstrated that DSAE-PBHL can subtract distributed representations of acoustic signals enabling the extraction of speaker-independent feature representation to a certain extent. 
The performance was quantitatively evaluated. 
, but depends on types of phonemes. The second experiment demonstrated that the combination of NPB-DAA and DSAE-PB outperformed the available unsupervised learning methods in phoneme and word discovery tasks with speech signals with Japanese vowel sequences from multiple speakers.

The future challenges are as follows:
The experiment was performed on vowel signals. However, applying NPB-DAA to more natural speech corpora is our future challenge. It would involve consonants, which exhibit more dynamic features than vowels. However, achieving unsupervised phoneme and word discovery from natural corpora including consonants and common vocabularies continues to be a challenging problem. Tada et al. applied NPB-DAA with a variety of feature extraction methods~\citep{YukiTada2017}. However, they obtained limited performance. Therefore, in this study, we focused on vowel data. Extending our studies to more natural spoken language is one of our intention.

Applying the method to larger corpora is another challenge. In this regard, the computational cost is high, and the method to address data from multiple speakers have been problematic. We consider our proposed method to have overcome one of these barriers. Recently, Ozaki et. al. reduced the computational cost of NPB-DAA significantly~\citep{RyoOzaki2018}. Therefore, we consider our contribution to be effective for further study of unsupervised phoneme and word discovery. 

This paper proposed DSAE-PBHL for proof-of-concept. DSAE-PBHL is regarded as a type of conditioned neural network.
Recently, the relationship between autoencoder and probabilistic generative model have been recognized via variational autoencoder~\citep{kingma2013auto}. 
From a broader perspective, our proposal is to use conditioned deep generative models to obtain disentangled representations to extract speaker-independent acoustic representations. In the field of speech synthesis, voice conversion methods using the generative adversarial network are studied~\citep{kameoka2018stargan}. We intend to explore the relationship between our proposal and such type of studies and integrate them in future studies.

In the current model, DSAE-PBHL and NPB-DAA are separately trained. However, as end-to-end learning in numerous deep learning-based models have indicated, the simultaneous optimization of feature extractor and post-processing is essential. We also intend to study the simultaneous optimization of representation learning and phoneme and word discovery in future.

\section*{Funding}
This work was supported by MEXT/JSPS KAKENHI Grant Number 16H06569 in \#4805 (Correspondence and Fusion of Artificial Intelligence and Brain Science) and 15H05319.

\section*{Data Availability Statement}
The datasets used for this study are available in our GitHub repository. 

Multi-speaker AIOI dataset  \url{https://github.com/EmergentSystemLabStudent/multi\_speaker\_aioi\_dataset}.

\bibliographystyle{frontiersinSCNS_ENG_HUMS} 
\bibliography{test}

\end{document}